\documentclass[twoside]{article}
\usepackage{fleqn,espcrc2}
\usepackage{graphicx}
\title{First results of the air shower experiment KASCADE}

\author{A.~Haungs\address{Institut f{\"u}r Kernphysik, 
         Forschungszentrum Karlsruhe, P.O. Box 3640, 
	 D--76021 Karlsruhe, Germany}\thanks{corresponding 
	 author, e-mail: haungs@ik3.fzk.de},
	T.~Antoni$^{\rm a}$, 
	W.D.~Apel$^{\rm a}$, 
        F.~Badea\address{Institute of Physics and Nuclear 
	Engineering, RO--7690 Bucharest, Romania},
	K.~Bekk$^{\rm a}$, 
	K. Bernl{\"o}hr$^{\rm a}$,
	H.~Bl\"umer$^{\rm a,d}$, 
	E.~Bollmann$^{\rm a}$, 
	H.~Bozdog$^{\rm b}$, 
	I.M.~Brancus$^{\rm b}$, 
	C.~B\"uttner$^{\rm a}$,
        A.~Chilingarian\address{Cosmic Ray Division, Yerevan Physics 
	                Institute, Yerevan 36, Armenia},
	K.~Daumiller\address{Institut f{\"u}r Experimentelle 
                          Kernphysik, Universit{\"a}t Karlsruhe, 
                          D--76021 Karlsruhe, Germany}, 
        P.~Doll$^{\rm a}$, 
	J.~Engler$^{\rm a}$, 
	F.~Fe{\ss}ler$^{\rm a}$, 
	H.J.~Gils$^{\rm a}$, 
	R.~Glasstetter$^{\rm d}$, 
	R.~Haeusler$^{\rm a}$,
	W.~Hafemann$^{\rm a}$, 
	D.~Heck$^{\rm a}$, 
	J.R.~H{\"o}randel$^{\rm d}$\thanks{now at: University of 
	                    Chicago, Chicago, IL 60637},
        T.~Holst$^{\rm a}$, 
	K.--H.~Kampert$^{\rm a,d}$, 
	H.~Keim$^{\rm a}$, 
	J.~Kempa\address{Department of Experimental Physics, 
	                 University of Lodz, PL--90236 Lodz, Poland},
	H.O.~Klages$^{\rm a}$, 
        J.~Knapp$^{\rm d}$\thanks{now at: University of Leeds, 
	                  Leeds LS2 9JT, U.K.},
	D.~Martello$^{\rm d}$,
        H.J.~Mathes$^{\rm a}$,
	P.~Matussek$^{\rm a}$, 
	H.J.~Mayer$^{\rm a}$, 
	J.~Milke$^{\rm a}$, 
	D.~M{\"u}hlenberg$^{\rm a}$, 
	J.~Oehlschl{\"a}ger$^{\rm a}$,
	M.~Petcu$^{\rm b}$, 
        H.~Rebel$^{\rm a}$, 
	M.~Risse$^{\rm a}$, 
	M.~Roth$^{\rm a}$, 
	G.~Schatz$^{\rm a}$, 
	F.K.~Schmidt$^{\rm a}$, 
        T.~Thouw$^{\rm a}$, 
	H.~Ulrich$^{\rm a}$, 
	A.~Vardanyan$^{\rm c}$,
	B.~Vulpescu$^{\rm b}$, 
	J.H.~Weber$^{\rm a}$, 
	J.~Wentz$^{\rm a}$, 
	T.~Wiegert$^{\rm a}$, 
        J.~Wochele$^{\rm a}$,
	J.~Zabierowski\address{Soltan Institute for Nuclear Studies,  
			 PL--90950 Lodz, Poland},
	S.~Zagromski$^{\rm a}$       }
       
\begin{document}

\begin{abstract}
The main goals of the KASCADE (KArlsruhe Shower Core and 
Array DEtector) experiment are the determination of the
energy spectrum and elemental composition of the charged
cosmic rays in the energy range around the knee at 
$\approx 5\,$PeV. Due to the large number of measured
observables per single shower a variety of different approaches
are applied to the data, preferably on an event-by-event basis.
First results are presented and the influence of
the high-energy interaction models underlying the 
analyses is discussed.  
\vspace{1pc}
\vspace{-1.pc}
\end{abstract}

\maketitle

\section{INTRODUCTION}
The air shower experiment 
KASCADE \cite{klages} aims at the investigation of the knee 
region of the charged cosmic rays. It is built up as
a multidetector setup for measuring simultaneously a large 
number of observables in the different particle (electromagnetic,
muonic and hadronic) components of the extended air shower (EAS). 
This enables to 
perform a multivariate multiparameter analysis for the
registered EAS on an event-by-event basis to account for the non
parametric, stochastic processes of the EAS development in the
atmosphere.
In parallel the KASCADE collaboration tries to improve the 
tools for the Monte Carlo simulations with the relevant
physics. The code CORSIKA \cite{cors} allows not only 
the detailed three dimensional simulation of the shower 
development in all particle components (including neutrinos) 
down to the observation level,
but it has been implemented several high-energy
interaction models. 
As the basic physics of these models in the relevant energy region
and in the extreme forward direction cannot be tested at 
present days' accelerators, the test of these models emerged as one of the
goals of the KASCADE experiment. 
The following overview is based on results presented at the 26$^{\rm
th}$ International Cosmic Ray Conference in Salt Lake City, Utah 
1999 \cite{slc}. 
  
\section{THE KASCADE EXPERIMENT}
The KASCADE array consists of 252 detector 
stations in a $200 \times 200\,$m$^2$ 
rectangular grid containing unshielded liquid 
scintillation detectors ($e/\gamma$-detectors) and below 10 cm 
lead and 4 cm steel plastic scintillators as muon-detectors.
The total sensitive areas are $490\,$m$^2$ for the $e/\gamma$-
and $622\,$m$^2$ for the muon-detectors. 
In the center of the array a hadron calorimeter 
($16 \times 20\,$m$^2$) is built up, consisting 
of more than 40,000 liquid ionisation chambers in 8 layers with 
a trigger layer consisting of 456 scintillation detectors 
in between.
Below the calorimeter a setup of position sensitive multiwire 
proportional chambers (MWPC) in two layers measures 
high-energy muons ($E_\mu > 2\,$GeV) of the EAS.  \\
For each single shower a large number of observables 
are reconstructed with small uncertainties.
For example, the errors for the so-called shower sizes, 
i.e. total numbers of electrons $N_e$ 
and number of muons in the range of the core distance 
$40-200\,$m $N_\mu^{tr}$, are smaller than 10$\%$. \\
The resulting frequency spectra of the sizes
(inclusive the spectra of the hadron number and muon density 
spectra at different core distances)
show kinks at same integral fluxes. This is a strong hint
for an astrophysical source of the knee phenomenon based on pure
experimental data, since same intensity of the flux corresponds
to equal primary energy. \\    
But for the reconstruction of the primary energy spectrum and
the chemical composition detailed Monte Carlo simulations are
indispensable due to the unknown initial parameters
and the large intrinsic fluctuations of the stochastic process of 
the shower development in the atmosphere. The usage of a larger
number of less correlated observables in a multivariate analysis
parallel to independent tests of the simulation models
tries to find the solution of this dilemma.  

\section{ANALYSES AND RESULTS}
\begin{figure}[htb]
 \begin{center}
\vspace*{-.cm}
 \includegraphics[width=16.5pc]{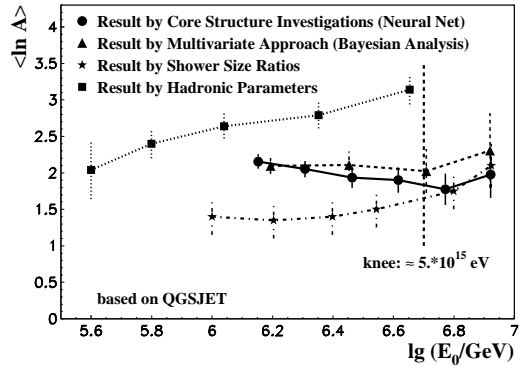}
 \end{center}
\vspace*{-1.3cm}
\caption{The chemical composition estimated with the KASCADE data, 
using different methods and observables from different particle
components.}
\label{fig:comp}
\vspace*{-.5cm}
\end{figure}

In the air shower simulation program CORSIKA several 
high-energy interaction models are embedded including 
VENUS, QGSJET and SIBYLL (Refs. see in \cite{cors}). 
The models are based on the Gribov-Regge-theory and QCD 
in accordance with accelerator data. 
Extrapolations for the EAS physics in the knee region
are necessary due to the high interaction energy and for
\begin{figure}[htb]
 \begin{center}
 \includegraphics[width=16pc]{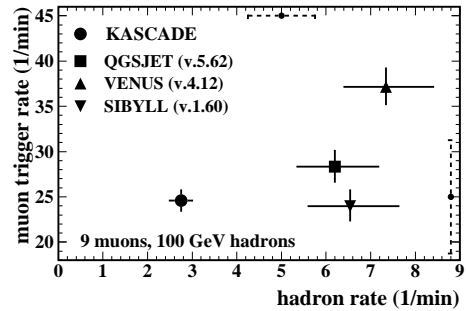}
 \end{center}
\vspace*{-1.25cm}
\caption{Comparison of simulated and measured integral muon 
trigger and hadron rates. Uncertainties of the elements' absolute
fluxes of the relevant energy range is indicated by dotted lines.}
\label{fig:rate}
\vspace*{-.5cm}
\end{figure}
the extreme forward direction. 
To compare KASCADE data with Monte Carlo expectations
a detector simulation by
GEANT is performed for each CORSIKA simulated shower. \\
One test is the comparison of simulated integral muon trigger and
hadron rates with the measurements. This test is sensitive 
to the energy spectrum of the hadrons which are produced 
in the forward direction at primary energies around 10 TeV, where the
chemical composition is roughly known (Fig.\ref{fig:rate}). 
For higher primary energies the hadronic part of the interaction
models are tested by comparisons of different hadronic 
observables in ranges of shower sizes \cite{antoni}. 
In general it is seen that the high-energy 
interaction models predict
a too large number of hadrons at sea level compared with the
measurements. \\
Nonparametric multivariate methods like ``Neural Networks''
or analyses based on the ``Bayesian decision rules'' 
are applied to the KASCADE data for the estimation of the 
energy and mass of the cosmic rays on an event-by-event basis. 
The necessary ``a-priori'' information in form of 
probability density distributions are won by detailed Monte Carlo 
simulations with large statistics. \\
For the energy reconstruction the shower sizes $N_e$ and 
$N_\mu^{tr}$ as parameters are used in a neural network analyses
(Fig.\ref{fig:spec}). 
A parametric approach to the same data 
leads to compatible results (Fig.\ref{fig:spec}):
\begin{figure}[htb]
\vspace*{0.1cm}
 \begin{center}
 \includegraphics[width=17pc]{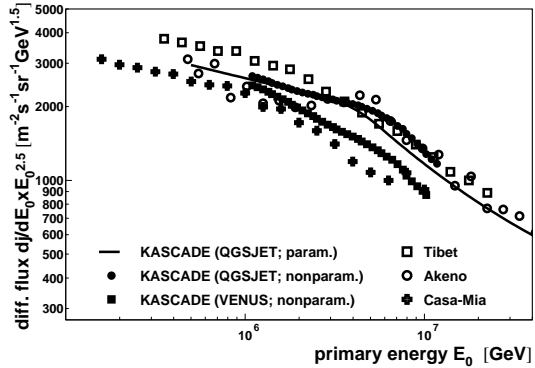}
 \end{center}
\vspace*{-1.2cm}
\caption{The primary cosmic ray energy spectrum from KASCADE
and other experiments. The spectral index changes from $\approx
\!-2.7$ to $\approx \!-3.1$ at the knee position of $\approx 5\cdot
10^6\,$GeV.}
\label{fig:spec}
\vspace*{-.5cm}
\end{figure}
here a simultaneous fit to the $N_e$ and $N_\mu^{tr}$
size spectra is performed.
The kernel function of this fit contains the size-energy 
correlations for two primary masses (proton and iron)
obtained by Monte Carlo simulations. \\
An analysis of the size-ratio $\lg (N_\mu^{tr})/ \lg (N_e)$ 
calculated for each single event leads to results of the 
elemental composition for different energy ranges 
(Fig.\ref{fig:comp}). 
The measured distribution of these ratios is assumed to be a 
superposition of simulated distributions for different primary 
masses. 
The large iron sampling calorimeter of KASCADE 
allows to investigate
the hadronic part of EAS in terms of the chemical
composition. For six different hadronic observables 
(won by spatial and energy distributions of the hadrons) 
the deviations of the mean values 
to expectations of pure proton and iron primaries 
in certain energy ranges are calculated. \\
Besides the use of global parameters like the shower sizes,  
sets of different parameters are used for neural network and 
Bayesian decision analyses. 
Examples of such observables are the number of reconstructed hadrons
in the calorimeter, their reconstructed energy sum, 
number of muons in the shower center, or
parameters obtained by a fractal analysis of the hit pattern of 
muons and secondaries at the MWPC.
The latter ones are sensitive to the structure of the shower core
which is mass sensitive due to different shower developments
of light and heavy particles in the atmosphere.
In Figure \ref{fig:comp} results of a Bayesian analyses 
and of a separate neural net analysis using the fractal parameters 
are shown. \\
As the tendency of the results of each described method is 
consistent with a heavier primary mass after the knee region, but
the absolute scale strongly depends on the particle component
of which the observables are constructed from, the syllogism
is that the balance of the energy and number of particles  
between the muonic, electromagnetic and hadronic part in the EAS 
differs for simulations and the real shower development.        
 
\section{CONCLUSIONS}
First results of the KASCADE experiment 
can be summarized by following statements:
All secondary particle components of the showers display a kink 
in the size spectra. 
This strongly supports an astrophysical origin of the ``knee'', 
rather than effects of the interaction of the primaries in the 
atmosphere.
The knee is sharper for the light primary component than
for the heavy one. 
This result follows from the measurement as an increasing average 
mass of the primary cosmic rays above the observed kink, together 
with the energy dependent mass classification of single air showers.
But none of the high-energy interaction models en vogue is 
able to fit the data of all observables consistently.

\end{document}